\documentclass[prd,nofootinbib,twocolumn]{revtex4}
\usepackage{graphicx, epsfig}
\usepackage{color}
\usepackage{mathrsfs}
\usepackage{bm}
\usepackage{amsmath,amssymb,yhmath}
\usepackage[caption=false]{subfig}
\usepackage{hyperref}
\usepackage[normalem]{ulem}

\usepackage{mathtools}
\usepackage{soul,xcolor}
\setul{1 ex}{1 ex}
\setstcolor{red}

\newcommand{\be}{\begin{equation}}
\newcommand{\ee}{\end{equation}}
\newcommand{\ba}{\begin{eqnarray}}
\newcommand{\ea}{\end{eqnarray}}

\newcommand{\half}{\frac{1}{2}}

\begin{document}
\title{Emergence of classical structures from the quantum vacuum}
\author{Mainak Mukhopadhyay$^{1}$}
\author{Tanmay Vachaspati$^{1}$}
\author{George Zahariade$^{1,2}$}
\affiliation{$^{1}$Physics Department, Arizona State University, Tempe, AZ 85287, USA. \\
}
\affiliation{$^{2}$Beyond Center for Fundamental Concepts in Science, Arizona State University, Tempe, Arizona 85287, USA} 

\begin{abstract}
\noindent
After a quantum phase transition the quantum vacuum can break up to form
classical topological defects. We examine this process for scalar field models
with $Z_2$ symmetry for different quench rates for the phase transition. 
We find that the number density of kinks at late times universally scales 
as $C m^{1/2} t^{-1/2}$ where 
$m$ is a mass scale in the model and $C\approx 0.22$; it does not depend on
the quench timescale in contrast to the Kibble-Zurek scaling for thermal phase
transitions. A subleading correction $\propto t^{-3/2}$ to the kink density depends 
on the details of the phase transition.
\end{abstract}

\maketitle

\section{Introduction}
\label{introduction}

Quantum field theories generally contain small quantum excitations around a true
vacuum that we call particles and large classical structures called solitons that
interpolate between different degenerate vacua. Often the solitons have a topological
character and are then also known as topological defects of which kinks, domain walls,
strings, and magnetic monopoles are all examples. 
As first proposed by Kibble~\cite{Kibble:1976sj}, these structures can be formed during 
a phase transition. A more quantitative estimate of their number density formed
after a {\it thermal} phase transition is given by the Kibble-Zurek 
proposal~\cite{Kibble:1980mv,Zurek:1985qw, Zurek:1993ek, Zurek:1996sj,Karra:1997he,
Vachaspati:2006zz,2007PhT....60i..47K}
that has been tested in various systems such as  liquid crystals~\cite{CHUANG1336,Bowick943},
superfluids~\cite{Hendry1994,PhysRevLett.81.3703,Ruutu1996,Bauerle1996,ELTSOV20051},
 superconductors~\cite{PhysRevLett.89.080603,PhysRevLett.84.4966,PhysRevLett.91.197001} 
and other systems involving liquid crystal light valves~\cite{PhysRevLett.83.5210} 
and ultra-cold quantum gases~\cite{Beugnon2017}, with conflicting conclusions. 
The Kibble-Zurek proposal
is based on imposing a physically motivated cutoff on the growing correlation length 
prior to the thermal phase transition and then matching the pre-phase transition
correlation length to the post-phase transition correlation length.
In this paper we will be concerned with a {\it quantum} phase transition and we
will solve for the {\it full} quantum dynamics relevant to defect formation.

To describe our approach we start with the $\lambda\phi^4$ model for a real, scalar field $\phi$,
\be
L = \half (\partial_\mu\phi)^2 - \half m_2(t) \phi^2 - \frac{\lambda}{4} \phi^4\,.
\label{lambdaphi4}
\ee
To study the production of kinks in this model we imagine that the mass parameter
$m_2(t)$ has an externally controlled time dependence,
\be
m_2(t) = - m^2 \tanh \left ( \frac{t}{\tau} \right )\,,
\label{m2}
\ee
where $\tau$ is the ``quench time scale''. For $t < 0$, the model has a unique
vacuum at $\phi =0$, while for $t > 0$, there are two degenerate vacua
$\phi = \pm m/\sqrt{\lambda} \equiv \pm \eta$.
In the $t \to \infty$ limit where $m_2 = -m^2$, the model has a 
double well potential and it admits
static classical kink and anti-kink solutions
\be
\phi_{\pm} (x) = \pm \eta \tanh \left ( \frac{m x}{\sqrt{2}}  \right )\,.
\label{kink}
\ee
These non-perturbative solutions interpolate between the two degenerate vacua of the model over a spatial width $\sim1/m$. They are topological defects characterized by a topological charge, positive for a kink and negative for an anti-kink. In fact the topological charge classifies kinks and anti-kinks according to the nature of the sign change ocurring in the field profile: negative to positive for a kink, and vice-versa for an anti-kink.

\begin{figure}
      \includegraphics[width=0.44\textwidth,angle=0]{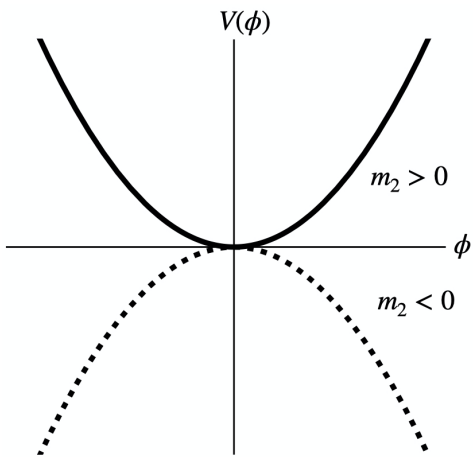}
  \caption{The field theory potential is an upright quadratic at early times
 when $m_2 > 0$ and becomes an inverted quadratic after the quench
 when $m_2 < 0$. The model has a $Z_2$ symmetry that is spontaneously
 broken and hence has kinks.
  }
  \label{potentialfigure}
\end{figure}

Since all we are interested in is changes in the sign of the field, 
we can simplify our model to eliminate the $\lambda\phi^4$ term in
\eqref{lambdaphi4} (see Fig.~\ref{potentialfigure}). Then the free field model
\be
L = \half (\partial_\mu\phi)^2 - \half m_2(t) \phi^2 
\label{model}
\ee
also has $Z_2$ symmetry that is spontaneously broken after the quench
and thus has topological kinks. Without the $\lambda\phi^4$ term, 
the kink height is not stabilized and becomes larger with time.
We expect that for a small enough value of $\lambda$, the interaction term does not significantly affect the kink number density. Hence the model in \eqref{model}
captures the essential physics of defect 
formation\footnote{One can think of this model as the $\lambda \to 0$, $\eta \to \infty$ limit 
of the $\lambda\phi^4$ model with $m=\eta\sqrt{\lambda}$ held fixed at some chosen 
finite value.}. It can serve as the zeroth order approximation in a perturbative
expansion in powers of $\lambda$.

In Sec.~\ref{qftsoln} we will solve the quantum field theory problem for the
model in \eqref{model} in the Schrodinger picture which is more convenient for
the kink number density calculation in Sec.~\ref{nksoln}. Our numerical results
for the kink number density and its dependence on various parameters are
given in Sec.~\ref{numsoln}. We conclude and qualitatively describe
expectations when $\lambda\ne 0$ in Sec.~\ref{conclusions}. Some technical
results are included in the Appendices.

\break

\section{Wavefunctional}
\label{qftsoln}

Setting $\lambda=0$, the problem now is that of a quantum field interacting with a classical background.
As discussed in~\cite{Vachaspati:2018hcu,Olle:2019skb,Mukhopadhyay:2019hnb}, the solution to the quantum problem can be written in
terms of the solution of a classical problem but in higher dimensions. More
specifically, let us discretize a compactified space of size $L$ by $N$ lattice points 
$i=1,\ldots,N$, with lattice spacing $a=L/N$. 
The resulting lattice-point-dependent Heisenberg picture field operators $\hat{\phi}_i$ describe a set of $N$, quadratically coupled, simple harmonic oscillators and we can write
\be
\hat{\phi}_i = Z_{ij}^* {\hat a}_j +  Z_{ij} {\hat a}_j^\dag\,,
\ee
where ${\hat a}_{j}$ and ${\hat a}_{j}^\dag$ are the annihilation and
creation operators associated with the quantum variable $\hat{\phi}_i$ at $t=-\infty$ when the potential is upright and time-independent.
The $N\times N$ complex matrix
$Z$ satisfies the classical equation,
\be
{\ddot Z} + \Omega_2 (t) Z = 0
\label{zeq}
\ee
where the matrix $\Omega_2$ is given by
\be
[\Omega_2]_{ij} = 
\begin{cases}
+{2}/{a^2}+m_2(t)\,,& i=j\\
-{1}/{a^2}\,,& i=j\pm1\ (\text{mod}\ N)\\
0\,,&\text{otherwise}\,.
\end{cases}
\ee
In our particular application, since $m_2(t)$ does not depend on the lattice point, 
all matrices are circulant ({\it i.e.} their $(i,j)$ element only depends on $i-j\ (\text{mod}\ N)$) and the problem is translationally invariant. We can thus diagonalize $Z$ to work in momentum space. This leads to
\be
Z_{jl}=  \frac{1}{\sqrt{N}} \sum_{n=1}^N c_n(t) e^{-i(j-l)2\pi n /N}\,,
\label{Zceq}
\ee
where the mode coefficients $c_n$ satisfy
\be
{\ddot c}_n + \left [ \frac{4}{a^2}\sin^2 \left (\frac{\pi n}{N} \right ) + m_2(t) \right ] c_n = 0\,.
\label{ckeq}
\ee
The quantum state (in Heisenberg picture) will be chosen to be the vacuum state long before the phase transition {\it i.e.} at $t=-\infty$ when $m_2(t)=+m^2$. In practice this is achieved by choosing an initial time $t_0\ll -\tau$ and setting up the following initial conditions for the mode coefficients:
\be
c_n (t_0) = 
\frac{-i}{\sqrt{2aN}} 
\left [ \frac{4}{a^2}\sin^2 \left ( \frac{\pi n}{N}\right ) + m_2(t_0) \right ]^{-1/4}\,,
\label{ckt0}
\ee
\be
{\dot c}_n (t_0) = \frac{1}{\sqrt{2aN}} 
\left [ \frac{4}{a^2}\sin^2 \left (\frac{\pi n}{N} \right ) + m_2(t_0) \right ]^{1/4}\,.
\label{dotckt0}
\ee

While it may be easier to solve for the quantum dynamics in momentum space by computing the $c_n$ mode coefficients, the kinks are defined as zeros of the field in physical space. It is therefore useful to determine the physical space wavefunctional $\Psi[\{\phi_i\},t]$ for the model in Eq.~\eqref{model} by solving the corresponding Schr\"odinger equation:
\be
i\frac{\partial \Psi}{\partial t}=-\frac{1}{2a}\sum_{i=1}^N\frac{\partial^2\Psi}{\partial\phi_i^2}+\frac{a}{2}\phi^T\Omega_2(t)\phi\,\Psi\,.
\ee
We find
\ba
\Psi [\{\phi_i\},t] &=& 
\left(\frac{a}{\pi}\right)^{N/4}{\rm det}\left(\Omega_2(t_0)\right)^{1/8}
\nonumber \\
&& \hskip -0.45 in
\times \exp \left [ -\half \int_0^t dt' {\rm Tr}\, M(t')
                                          + \frac{ia}{2} \phi^T M(t)\, \phi \right ],
                                          \label{Psisoln}
\ea
where $M \equiv {\dot Z}Z^{-1}$ and we have introduced the column vector $\phi$ such that $\phi^T \equiv (\phi_1,\dots,\phi_N)$. Using the constraint $Z^\dag\dot{Z}-\dot{Z}^\dag Z=i/a$~\cite{Vachaspati:2018hcu}, this gives a probability distribution for the field,
\be
P[\{\phi_i\},t] = \frac{1}{\sqrt{{\rm det}(2\pi K})} e^{- \phi^T K^{-1} \phi / 2}\,,
\label{pdist}
\ee
where $K=ZZ^\dag$ is the covariance matrix of the field $\phi$. Notice that $K$ is real and symmetric~\cite{Vachaspati:2018hcu} (as can be verified by using~\eqref{zeq} and 
the initial conditions).

\section{Number density of kinks}
\label{nksoln}

The number density of zeros ($n_Z$) is obtained by counting the number of sign changes of $\phi$. To compute an explicit formula, we first define the quantum operator
\be
\hat{n}_Z \equiv \frac{1}{L} \sum_{j=1}^N \frac{1}{4} \left [
\text{sgn}\left(\hat{\phi}_j\right) - \text{sgn}\left(\hat{\phi}_{j+1}\right) \right ]^2\,, 
\ee
where, because of the periodicity of the lattice, $\hat{\phi}_{N+1}=\hat{\phi}_1$. The number density of zeros is then simply given by the quantum average of this operator. After using translational invariance and in particular the fact that $K^{-1}$ is circulant  it reads (see Appendix~\ref{circulant} for details),
\be
n_Z= \langle\hat{n}_Z\rangle = \frac{N}{2L} \left [ 1 -
 \left \langle \text{sgn}\left(\hat{\phi}_1\hat{\phi}_2\right)  \right \rangle \right ]\,.
\label{nZ}
\ee

The expectation value in \eqref{nZ} can now be written as
\ba
\left \langle \text{sgn}\left(\hat{\phi}_1\hat{\phi}_2\right)  \right \rangle
&=&\frac{1}{\sqrt{{\rm det}(2\pi K})}
\nonumber \\
&& \hskip -0.9 in
\times
 \sum_{\rm quads.} \int d\phi_1\ldots d\phi_N\, {\rm sgn}(\phi_1\phi_2)\,  e^{- \phi^T K^{-1} \phi / 2},
\label{int1}
\ea
where the sum is over the four quadrants in the $(\phi_1,\phi_2)$ plane. The
coefficient, ${\rm sgn}(\phi_1\phi_2)$ is $+1$ for quadrants with 
$\phi_1\phi_2 > 0$ and $-1$ for quadrants with $\phi_1\phi_2 < 0$.

Consider the integral in the first quadrant,
\be
I_1 = \int_0^\infty d\phi_1 \int_0^\infty d\phi_2 \int_{-\infty}^\infty d\phi_3 \ldots d\phi_N 
\  e^{- \phi^\dag K^{-1}  \phi /2 }\,.
\ee
To perform the integration, we write
\be
K^{-1}  = \begin{pmatrix} A & B \\ B^T & C \end{pmatrix}\,,
\ee
where $A$ a $2\times 2$ matrix, $B$ a $2\times (N-2)$ matrix, and $C$ an
$(N-2)\times (N-2)$ matrix. (The matrices $A$ and $C$ are symmetric.)
Next, we also introduce the vectors $\chi$ and $\xi$ such that $\chi^T=(\phi_1,\phi_2)$ and $\xi^T=(\phi_3,\dots,\phi_N)$. Then, 
\ba
\phi^T K^{-1} \phi 
&& = (\xi + C^{-1} B^T \chi  )^T C (\xi + C^{-1} B^T \chi )  \nonumber \\
&& \hskip 0.6 in +\chi^T (A-BC^{-1}B^T) \chi \,,
\label{wholesq}
\ea
and we can perform the integration over $\xi$ first using
\be
\int d^{N-2}\xi ~ e^{- (\xi + C^{-1} B^T \chi  )^T C (\xi + C^{-1} B^T \chi ) /2} 
= 
\frac{(2\pi)^{(N-2)/2}}{ \sqrt{{\rm det}(C)}}\,. \nonumber
\ee
As detailed in Appendix~\ref{integrals} we obtain,
\ba
I_1 &=& 
\frac{(2\pi)^{(N-2)/2}}{ \sqrt{{\rm det}(C)}}
\int_0^\infty d\phi_1 \int_0^\infty d\phi_2 \,
e^{- \chi^T A'  \chi /2 } \nonumber  \\
&&\hskip -0.5 in = 
\frac{(2\pi)^{(N-2)/2}}
{\sqrt{{\rm det}(C) {\rm det}(A')}} 
\left [ \frac{\pi}{2} - \tan^{-1} \left (
\frac{A'_{12}} {\sqrt{{\rm det}(A')} }\right ) \right ],
\ea
where $A'$, the Schur complement of $C$, is defined by $A'\equiv A-B C^{-1} B^T$.

The integral over the third quadrant in the $\phi_1\phi_2$ plane, $I_3$, 
is seen to be equal to $I_1$ after the change of variables $\chi\rightarrow-\chi$.
The integrals over the second and fourth quadrants, $I_2$ and $I_4$, are similarly equal,
and are related to $I_1$. To see this we note
that the integral over the second quadrant reduces to the one over the first quadrant
by the transformation $\phi_1 \to -\phi_1$.
This transformation is alternatively implemented by changing 
$A'_{12}$ to $-A'_{12}$ and not changing anything else. Hence the
integral over the second quadrant is simply
\ba
I_2  = 
\frac{ (2\pi )^{(N-2)/2} }
{\sqrt{{\rm det}(C) {\rm det}(A')}} 
\left [ \frac{\pi}{2} + \tan^{-1} \left (
\frac{A'_{12}} {\sqrt{{\rm det}(A')} }\right ) \right ]\,.
\ea
Putting together the
contributions of all the four quadrants, we get
\be
\left \langle \text{sgn}\left(\hat{\phi}_1\hat{\phi}_2\right)  \right \rangle =
- \frac{2}{\pi} 
\tan^{-1} \left ( \frac{A'_{12}} {\sqrt{{\rm det}(A')} }\right )\,,
\ee
where we made use of the fact that $A'$ is the Schur complement of $C$ which implies
\be
{\rm det}(A') {\rm det}(C)={\rm det}(K^{-1})=\frac{1}{{\rm det}(K)}\nonumber\,.
\ee
This in turn gives us the number density of zeros as
\be
n_Z = \frac{N}{2L} \biggl [ 1 + 
 \frac{2}{\pi}  \tan^{-1} \left ( \frac{A'_{12}} {\sqrt{{\rm det}(A')} }\right ) \biggr ]\,.
\label{nKwithmatrices}
\ee
We can make this formula more explicit by noticing that $(A')^{-1}$ is equal to the 
$2\times 2$ upper left block of the matrix $K$. This can be seen via a block LDU 
decomposition of $K^{-1}$ 
as described in Appendix~\ref{block}. Using~\eqref{Zceq} we can then write
\be
(A')^{-1} = K_{2\times 2} = \alpha {\bf 1} + \beta \sigma_x\,,
\ee
where $\sigma_x$ is the first Pauli spin matrix and
\be
\alpha = \sum_{n=1}^N |c_n|^2 , \ \ 
\beta = \sum_{n=1}^N |c_n|^2 \cos (2\pi n/N)\,.
\ee
Therefore
\be
A' = \frac{1}{\alpha^2-\beta^2} \left ( \alpha {\bf 1} - \beta \sigma_x \right )\,,
\ee
and 
\be
{\rm det}(A') = \frac{1}{\alpha^2-\beta^2}, \ \ \ 
A'_{12} = \frac{-\beta}{\alpha^2-\beta^2}\,.
\label{a'props}
\ee

We now have all the pieces needed to evaluate the number density of zeros in
\eqref{nKwithmatrices} which can be written as
\be
n_Z = \frac{N}{2L} \left [ 1 - \frac{2}{\pi} \theta \right ]\,,
\ee
where
\be
\sin\theta \equiv \frac{\beta}{\alpha} 
=\frac{\sum_{n=1}^N |c_n|^2 \cos (2\pi n/N) }{ \sum_{n=1}^N |c_n|^2 }\,.
\label{sintheta}
\ee

Not all field zeros, however, correspond to kinks. Some of
the zeros simply correspond to field oscillations that come in and out of existence.
They are spurious or virtual kinks and we eliminate them from our counting by
restricting the summations in \eqref{sintheta} to modes that aren't 
oscillating\footnote{This is similar to the situation in inflationary cosmology
where only non-oscillating super-horizon modes lead to density perturbations.}.
So the number density of kinks, $n_K$, is
\be
n_K = \frac{N}{2L} \left [ 1 - \frac{2}{\pi} \sin^{-1} \left (
\frac{\sum_{|n| \leq n_c} |c_n|^2 \cos (2\pi n/N) }{ \sum_{|n| \leq n_c} |c_n|^2 }
\right ) \right ]\,,
\label{nK}
\ee
where, as seen in \eqref{ckeq}, the time-dependent cut-off mode for $t > 0$ is
defined by
\be
\sin \left ( \frac{\pi n_c(t)}{N} \right ) = \frac{a}{2} \sqrt{|m_2(t)|}\,,
\ee
and $c_{-n}\equiv c_{N-n}$ for $0 \leq n \leq N-1$.

\section{Numerical results}
\label{numsoln}

\begin{figure}[ht]
      \includegraphics[width=0.45\textwidth,angle=0]{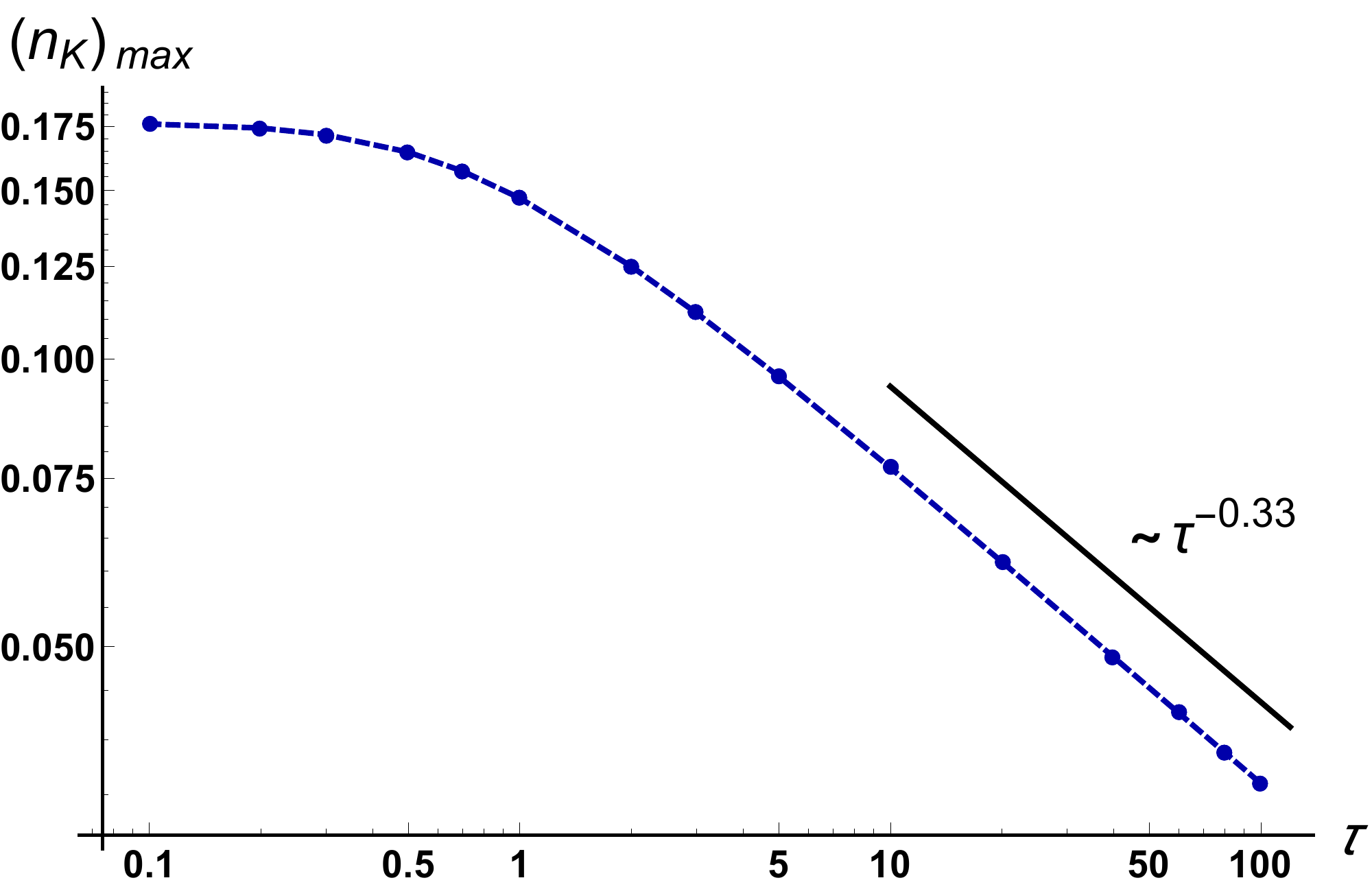}
  \caption{Log-Log plot of the maximum average kink density $(n_K)_{\rm max}$ vs. $\tau$. 
  For larger values of $\tau$ the power law manifested is $\sim \tau^{-0.33}$.}
  \label{maxnkvstau}
\end{figure}

\begin{figure}
      \includegraphics[width=0.45\textwidth,angle=0]{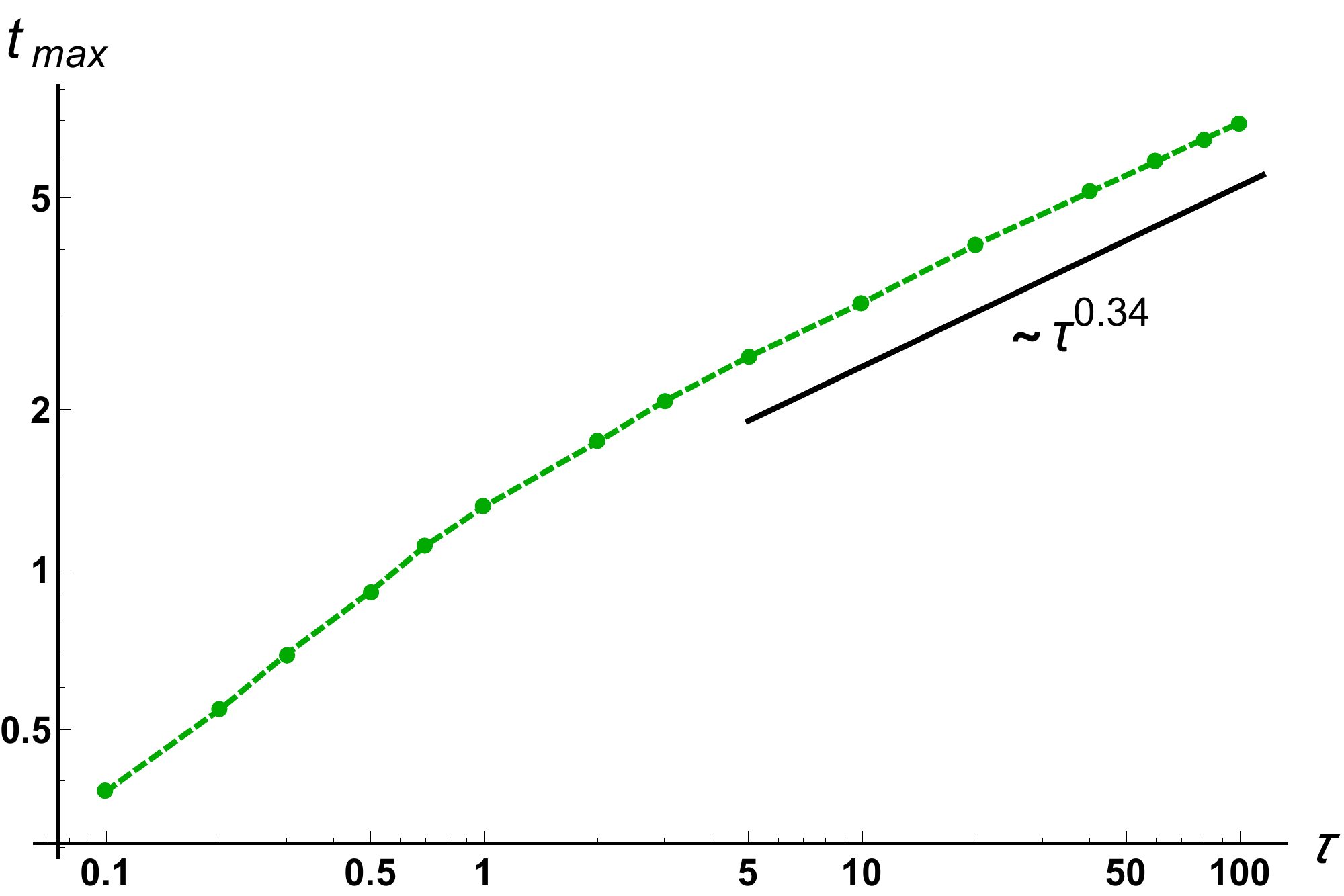}
  \caption{Log-Log plot of the time at which maximum average kink density $(n_K)_{\rm max}$ 
  occurs vs. $\tau$. For larger values of $\tau$ the power law exhibited is $\sim \tau^{0.34}$.}
  \label{maxnkt}
\end{figure}

We can now numerically solve \eqref{ckeq} and use \eqref{nK} to obtain the number density
of kinks as a function of time. 
The only scale in the problem is the mass so we work in units of $1/m$ by setting $m=1$. 
We also choose $L= 6400$ and $N= 12800$,
which are both large enough to accurately describe 
the continuum, infinite space limit of the discretized model. Notice that, thanks to the physical 
cutoff we placed on the mode sums, there are no UV divergences.\\
\begin{figure}[ht]
      \includegraphics[width=0.45\textwidth,angle=0]{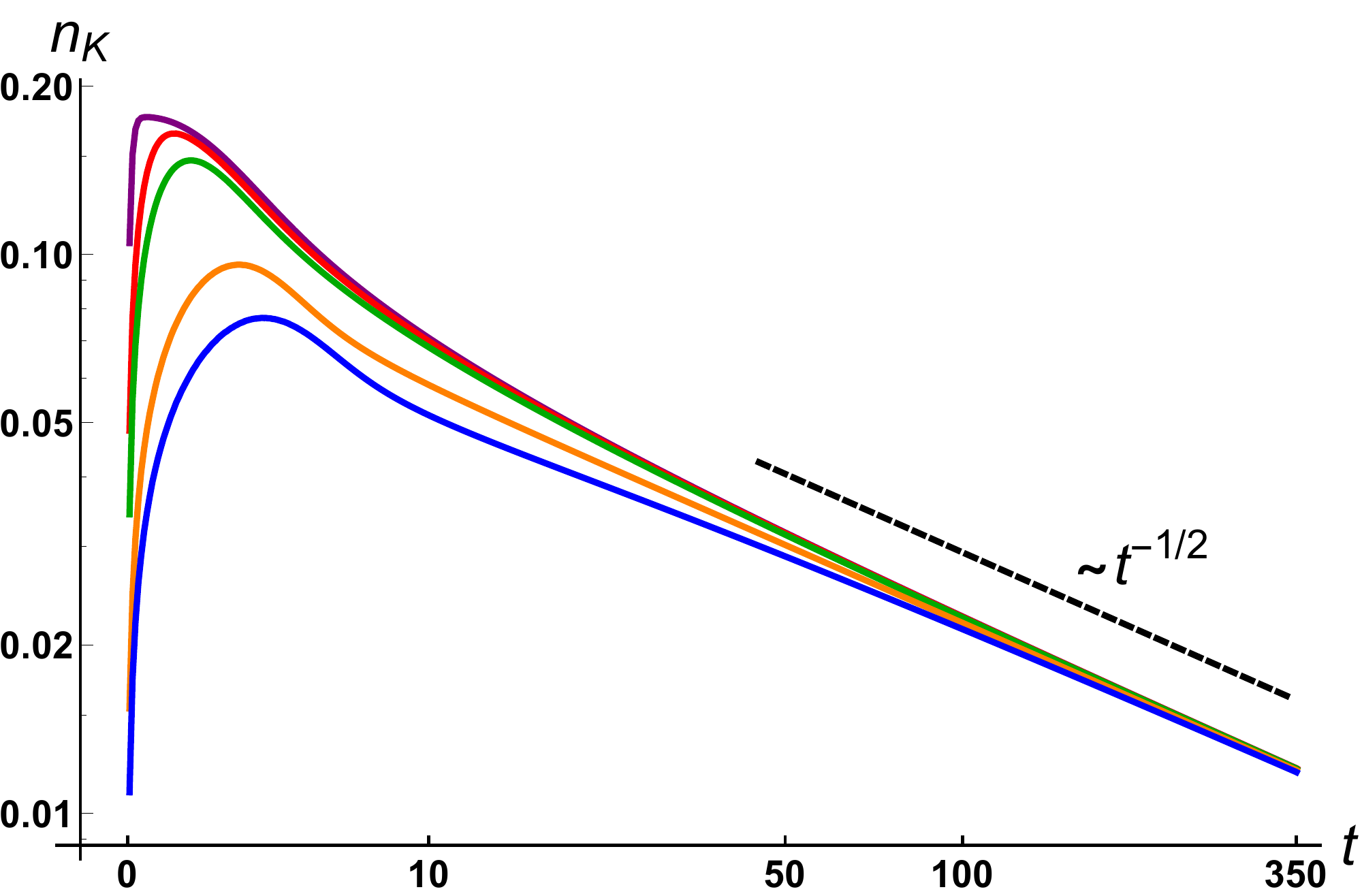}
  \caption{Log-log plot of $\langle n_K \rangle$ versus time for $\tau$ =0.1 (Purple, topmost curve), 0.5 (Red), 1.0 (Green), 5.0 (Orange), 10.0 (Blue). The black dashed line shows the exhibited power law at late times, \emph{i.e.} $t^{-1/2}$.}
  \label{nKvstloglog}
\end{figure}
\begin{figure}[ht]
      \includegraphics[width=0.45\textwidth,angle=0]{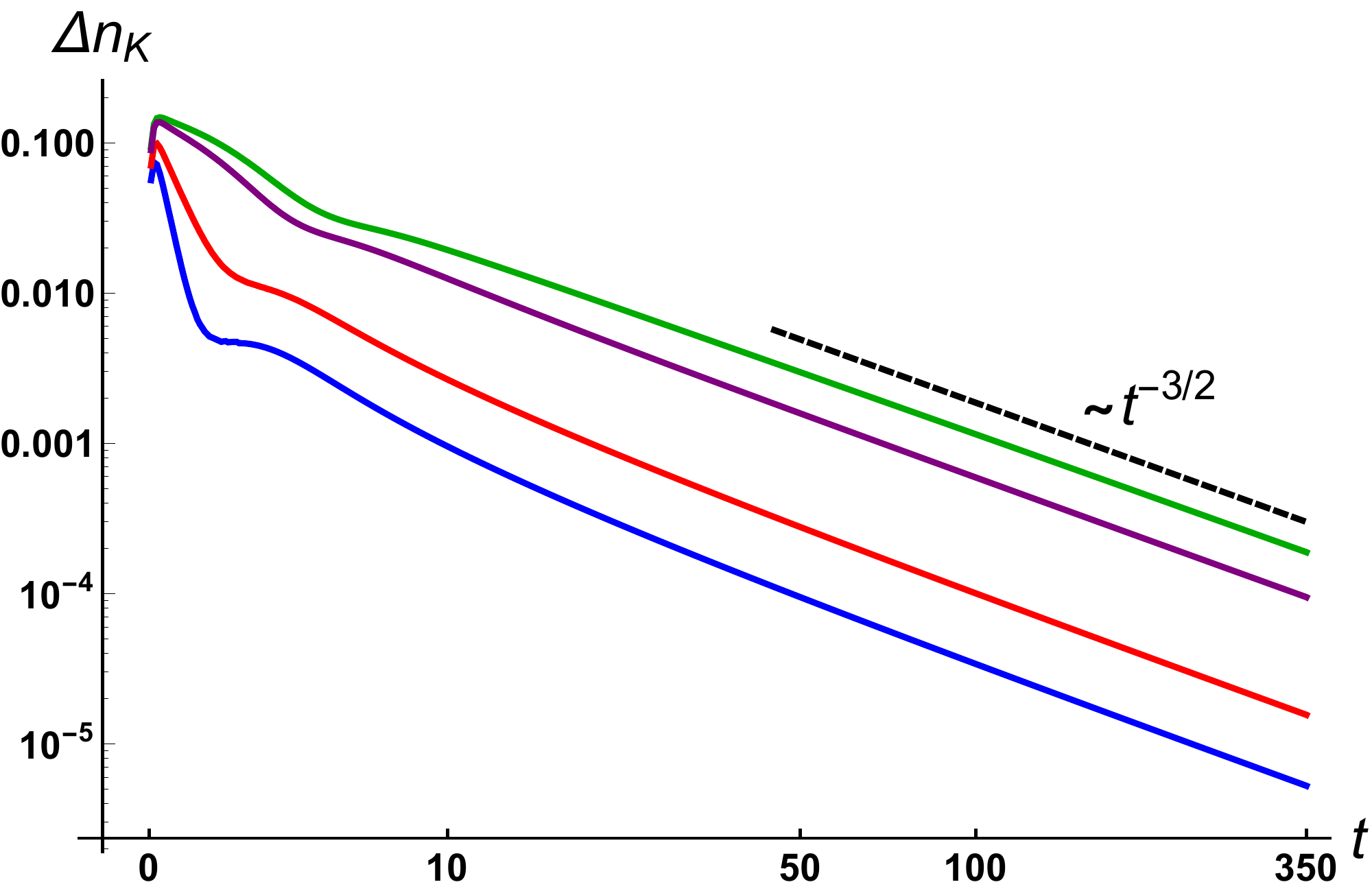}
  \caption{Differences between the average kink density for different values of $\tau$,
$n_K(t,\tau_1) - n_K(t,\tau_2)$ vs. time, for
$\tau_1 = 0.1$; $\tau_2 = $ 0.5 (Blue, bottom-most curve), 1.0 (Red), 5.0 (Purple), 10.0 (Green). 
The black dashed line shows the exhibited power law, \emph{i.e.} $t^{-3/2}$.}
  \label{difference}
\end{figure}
In Fig.~\ref{nKvstloglog} we show our results for 
several different values of the quench parameter $\tau$. The remarkable
feature of this plot is that all the curves have the same late time  behavior which
we can determine 
to be a $t^{-1/2}$ power law. In fact, we can take differences for different values of $\tau$, 
$\Delta n_K(t,\tau_1,\tau_2) \equiv n_K(t,\tau_1) - n_K(t,\tau_2)$ (see Fig. \ref{difference}), and 
these follow a $t^{-3/2}$ power law.
Thus at late times we can write
\be
n_K (t) = C \sqrt{\frac{m}{t}} \, + \, \mathcal{O}\left ( t^{-3/2} \right )\,,
\label{nKresult}
\ee
where we get $C \approx 0.22$ from our numerical solution. Note that
$C$ is independent of $\tau$.
At early times ({\it i.e.} immediately after the phase transition), $n_K$ increases from 0 to a 
maximum value $(n_K)_{\rm max}$ within a time $t_{\rm max}$, before decreasing again. 
This is to be expected: the phase transition triggers the creation of kinks with randomly 
distributed positions and velocities, which later start annihilating with each other. In 
Figs.~\ref{maxnkvstau} and~\ref{maxnkt} we plot $(n_K)_{\rm max}$ and $t_{\rm max}$ 
respectively, as a function of the quench parameter $\tau$. This  confirms the intuitive 
expectation according to which the faster the phase transition (smaller $\tau$), the more 
kinks are produced and the quicker they start annihilating.

\section{Conclusions}
\label{conclusions}

Our key result is Eq.~\eqref{nKresult}. It tells us how the quantum vacuum
breaks up into classical solitons after a quench. Further, it shows that the
result at late times is universal and does not depend on the quench timescale
even though there is some dependence closer to the time of the phase transition.
In the Kibble-Zurek proposal for kinks produced 
during a {\it thermal} phase transition, the kink density immediately after the phase 
transition depends on the quench timescale and is proportional to $\tau^{-1/4}$
in certain systems~\cite{Laguna:1996pv}.
This is to be contrasted with the $\tau^{-1/3}$ fall off in Fig.~\ref{maxnkvstau}.

We have also cross-checked our results by numerically estimating $n_K$ as the 
inverse of the correlation length that is extracted from the covariance matrix $K$,
as well as analytically by studying the limiting case of a sudden phase transition
with $\tau=0$. We plan to present these cross-checks in a separate publication~\cite{progress}.
Similar analyses for a sudden ($\tau=0$) thermal quench have been done 
in~\cite{PhysRevE.48.767,Boyanovsky:1999wd} by different techniques;
a $t^{-1/2}$ scaling of defect density was also observed
but the dependence on quench timescale was not studied.

In our approach we have taken vanishing interaction strength, $\lambda=0$, and 
it is of interest to ask how the results might change if $\lambda$ is different from
zero. As we have noted in the introduction, our result for the number density of kinks 
can be thought of as the ${\cal O}(\lambda^0)$ term in a perturbation expansion in $\lambda$.
It should be possible to compute higher order corrections in $\lambda$ using perturbation 
theory. In particular,
there will be $\lambda$ dependent corrections to our wavefunctional in \eqref{Psisoln}.
However the corrections to the number density of kinks also depend on the 
parameter $\tau$ that determines how quickly the potential for the field $\phi$
changes. If $\tau$ is very small, the wavefunctional can be computed in the
``sudden approximation''~\cite{progress}
which will be valid if 
the time-scale for changes in the potential is much shorter than the
time-scale set by the interaction term.


Our analysis also suggests that the interaction between kinks is not important for sudden phase transitions. As can be seen in Fig.~\ref{maxnkvstau}, the maximum kink number density for $\tau=0$ is $(n_{K})_{\rm max}
\approx0.175$
, and the average 
kink separation, $n_K^{-1}$, is greater than at least $\sim 6$ times the width of the kink.
Since the attractive kink-anti-kink force decreases exponentially with distance in 
$1+1$ dimensions, the effect of inter-kink forces can be consistently disregarded.

The analysis we have done in this paper can be generalized to higher dimensions
to discuss global vortex formation in two spatial dimensions and global monopole
formation in three spatial dimensions, since the models for these can be truncated
to free fields in time dependent backgrounds~\cite{progress}. The introduction of gauge 
fields, however, will lead to new interactions that will be more difficult to analyze.

Finally it is worth mentioning that there is a deeper foundational question in this
problem that we have studied. Our initial state is a translationally invariant quantum
vacuum, while the final state involves classical kinks with definite positions and
velocities. The translational symmetry is preserved when averages are taken over 
an ensemble of kink realizations, but each realization of the kinks breaks translational 
symmetry. As in Schrodinger's cat, the classical kinks materialize and break translational
symmetry only when there is a detector that detects them.

\acknowledgements
We thank Dan Boyanovsky for comments. MM is supported by the National Science Foundation grant numbers PHY-1613708 and PHY-2012195. TV is supported by the U.S.  Department of Energy, Office of High Energy Physics, under Award No.~DE-SC0019470 at Arizona State University. GZ is supported by {\it Moogsoft} and the Foundational Questions Institute (FQXi).

\appendix

\section{Some Properties of Circulant Matrices}
\label{circulant}

\noindent A generic $N\times N$ matrix $A=(a_{ij})$ is circulant when its entries $a_{ij}$ only depend on $j-i\ ({\rm mod}\ N)$. In other words $a_{ij}\equiv a_{j-i\ ({\rm mod}\ N)}$. It is useful to introduce the permutation matrix 
\be
P_{ij}=
\begin{cases}
1\,,&j=i+1\, (\text{mod}\ N)\\
0\,,&\text{otherwise}\,,
\end{cases}
\ee
verifying $P^{-1}=P^T=P^{N-1}$. The matrix $A$ is seen to be a degree $N-1$ polynomial in $P$:
\be
A=a_0 I + a_1 P + a_2 P^2 +\dots +a_{N-1}P^{N-1}\ .
\ee
This immediately implies that sums, products, transposes and Hermitian conjugates of circulant matrices are also circulant (and that circulant matrices commute with each other). An additional consequence is that bilinears involving circulant matrices are unchanged under circular permutations of vector elements. Indeed, given a positive integer $n\leq N-1$ and two vectors $X$ and $Y$, we have
\be
(P^nX)^TA(P^nY)=X^TP^{-n}AP^nY=X^TAY\ .
\ee
Moreover the circulant matrix $A$ can be diagonalized via the unitary discrete Fourier transform matrix $F=(f_{jk})$ where $f_{jk}\equiv\frac{1}{\sqrt{N}}e^{-i2\pi jk/N}$:
\be
A= F^{-1}D F = F^{\dag}DF\ .
\ee
This shows that if a circulant matrix is invertible, then its inverse is also circulant. All these properties are often used in the main text, and in particular when using the fact that $\left\langle {\rm sgn} \left(\hat{\phi}_i\hat{\phi}_{i+1}\right)\right\rangle= \left\langle {\rm sgn} \left(\hat{\phi}_1\hat{\phi}_{2}\right)\right\rangle$ in the lead-up to Eq. (16).

\section{Two-dimensional Gaussian integral over the First Quadrant}
\label{integrals}

\noindent The derivation of the two-dimensional Gaussian integral over the first quadrant $x,y\geq 0$ follows from the change of variables $x=sy$:
\ba
&&\hspace{-1cm}\int_0^\infty dx \int_{0}^\infty dy\, e^{-\frac{1}{2}(ax^2+2bxy +cy^2)}\nonumber\\
&=&\int_0^{\infty} ds \int_{0}^{\infty} dy\,y\, e^{-\frac{y^2}{2}(as^2+2bs +c)}\nonumber\\
&=&\int_0^\infty \frac{ds}{as^2+2bs+c}\nonumber\\
&=&\frac{1}{\sqrt{ac-b^2}}\left[\frac{\pi}{2}-\tan^{-1}\left(\frac{b}{\sqrt{ac-b^2}}\right)\right]\ .
\ea
Here we have assumed that $a$, $b$ and $c$ are real numbers such that $a>0$ and $ac-b^2>0$. This ensures that all expressions in the above equations are well posed. This identity is used in the main text in the lead-up to Eq. (21).

\section{Some Properties of Block Matrices}
\label{block}

\noindent Consider a square matrix $M$ partitioned into four blocks of arbitrary size as follows:
\be
M=
\begin{pmatrix}
A & B \\
C & D
\end{pmatrix}\ .
\ee
Performing the analog of an LDU decomposition on this block matrix, we can write
\ba
M&=&
\begin{pmatrix}
A & B \\
C & D
\end{pmatrix}
 \\ \nonumber &=& 
\begin{pmatrix}
I & BD^{-1} \\
0 & I
\end{pmatrix}
\begin{pmatrix}
A-BD^{-1}C & 0 \\
0 & D
\end{pmatrix}
\begin{pmatrix}
I & 0 \\
D^{-1}C & I
\end{pmatrix}\,.
\ea
\\
Here we have assumed that $A$ and $D$ are square invertible matrices. Note that the identity matrices can have different sizes. This expression immediately implies that
\be
{\rm det} M = {\rm det} \left(A-BD^{-1}C\right)\,{\rm det} D\ .
\ee
The matrix $A-BD^{-1}C$ is called the Schur complement of $D$. This formula is used in the main text  in the lead-up to Eq. (23). Assuming invertibility of the Schur complement we can also write
\begin{widetext}
\ba
M^{-1}&=& 
\begin{pmatrix}
I & 0 \\
-D^{-1}C & I
\end{pmatrix}
\begin{pmatrix}
\left(A-BD^{-1}C\right)^{-1} & 0 \\
0 & D^{-1}
\end{pmatrix}
\begin{pmatrix}
I & -BD^{-1} \\
0 & I
\end{pmatrix}\nonumber\\
&=&
\begin{pmatrix}
\left(A-BD^{-1}C\right)^{-1} & -\left(A-BD^{-1}C\right)^{-1}BD^{-1} \\
-D^{-1}C\left(A-BD^{-1}C\right)^{-1} & D^{-1}+D^{-1}C\left(A-BD^{-1}C\right)^{-1}BD^{-1} 
\end{pmatrix}\ ,
\ea
\end{widetext}
which allows one to directly read off the inverse of the Schur complement of $D$ on the block decomposition of $M^{-1}$. This identity is used in the main text in the lead-up to Eq. (25).
\break

\bibstyle{aps}
\bibliography{paper}

\end{document}